\documentclass[prb,twocolumn,showpacs,amsmath,amssymb,superscriptaddress,preprintnumbers]{revtex4}
\usepackage{graphicx,ulem}
\usepackage{enumerate}
\newcommand \beq{\begin{eqnarray}}
\newcommand \eeq{\end{eqnarray}}

\newcommand{\bfr}{\mathbf{r}}
\newcommand{\bfR}{\mathbf{R}}
\newcommand{\bfk}{\mathbf{k}}
\newcommand{\bfK}{\mathbf{K}}
\newcommand{\bfs}{\mathbf{s}}

\newcommand{\vf}{v_{_F}}

\newcommand{\im}{\mathrm{Im}}

\begin{document}

\preprint{RIKEN-QHP-78, RIKEN-MP-68}

\title{Phase structure of 2-dimensional topological insulators by lattice strong coupling expansion}
\author{Yasufumi Araki}
\affiliation{Quantum Hadron Physics Laboratory, RIKEN Nishina Center, Saitama 351-0198, Japan}
\affiliation{Department of Physics, The University of Tokyo, Tokyo 113-0033, Japan}
\author{Taro Kimura}
\affiliation{Mathematical Physics Laboratory, RIKEN Nishina Center, Saitama 351-0198, Japan}

\begin{abstract}
The phase structure of 2-dimensional topological insulators
under a sufficiently strong electron-electron interaction is investigated.
The effective theory is constructed by extending the idea of the Kane-Mel\'e model
on the graphenelike honeycomb lattice,
in terms of U(1) lattice gauge theory (quantum electrodynamics, QED).
We analyze the phase structure by the
techniques of strong coupling expansion of lattice gauge theory.
As a result, we find
that the topological phase structure of the system is modified by the electron-electron interaction.
There evolves a new phase with the antiferromagnetism not parallel to the direction pointed by the spin-orbit coupling,
in between the conventional and the topological insulator phases.
{We} also discuss the physical implication of the new phase structure 
found here, in analogy
to the {parity-broken} phase in lattice quantum chromodynamics (QCD), known as ``Aoki phase''.
\end{abstract}
\pacs{73.22.Pr,03.65.Vf,11.15.Ha,11.15.Me}
\maketitle

\section{Introduction}
Topological insulators have recently attracting a great interest in the field of materials physics \cite{Hasan-Kane,Qi-Zhang}.
They are characterized by the gapless modes localized on the surfaces or edges of the system,
while the bulk spectrum is separated by a finite bandgap.
The existence of such gapless modes is ensured by a nontrivial topological invariant (number)
defined by the electron ground state.
The gapless boundary modes have the properties of massless Dirac fermions,
which are topologically protected under any consistent perturbation or disorder {with the symmetry}.

The 2-dimensional quantum spin Hall (QSH) insulator is one of the examples for topological insulators,
where the shift of the topology is given by the spin-orbit coupling acting on the electrons.
It was first observed in HgTe quantum wells experimentally in 2007 \cite{Konig_2007},
{which agrees with the theoretical model proposed previously \cite{Bernevig-Hughes-Zhang}}.
The band structure of QSH insulators is effectively described by the Kane--Mel\'e model,
which is based on the effective theory of graphene on the honeycomb lattice \cite{Kane-Mele}.
The system possesses a finite {$Z_2$ topological number},
which gives rise to QSH effect even in the absence of external magnetic field,
related by the so-called Thouless--Kohmoto--Nightingale--den Nijs (TKNN) formula \cite{TKNN}.
Its topological phase structure is characterized by the competition between
the topological gap from the spin-orbit interaction and the non-topological gap from some other symmetry breaking effects.
Change of the topology occurs at the phase boundary, where one of the valleys loses its bandgap.

The effect of electron correlation in such an electronic system has always been an important problem.
Even in non-topological Dirac fermion systems, such as graphene,
it has been proposed that a sufficiently strong electron-electron interaction can lead to
a spontaneous breaking of some symmetries of the system and a dynamical generation of bandgap \cite{Physics_2009}.
In some of the previous studies,
the idea of quantum electrodynamics (QED),
such as Schwinger--Dyson equation \cite{Gorbar_2002}, large-$N$ expansion \cite{Herbut_2006,Son_2007},
exact renormalization group analysis \cite{ERG-cont,ERG-honeycomb},
Monte Carlo simulation \cite{Drut,Shintani,Buividovich} and strong coupling expansion \cite{Araki_2010,Araki-honeycomb} of lattice gauge theory,
has been applied to study the effect of electron-electron interaction in graphene {(or graphenelike)} system.
It has been predicted that the system can show a rich phase structure
depending on the pattern of symmetry breaking.

In this paper, we study the effect of a sufficiently strong electron-electron interaction
on the topological phase structure of 2D QSH (topological) insulators.
We extend the idea of the strong coupling expansion analysis on the lattice gauge theory of graphene
by adding the effect of spin-orbit interaction like the Kane--Mel\'e model.
In the strong coupling limit of the electron-electron interaction,
there appears an antiferromagnetic (AF) order spontaneously,
and we observe the behavior of the order parameter by varying the amplitude of topological and non-topological gaps.
As a result, we find that the topological phase structure of the system is modified
from that of the noninteracting system.
A new phase, which we call here ``tilted AF'' phase,
evolves around the phase boundary between the topological and non-topological insulator phases,
where the direction of the antiferromagnetic order is different from that pointed by the spin-orbit interaction
in the SU(2) spin space.
In such a phase, we expect that the system can possess a gapless Nambu--Goldstone mode,
in contrast to the conventional topological and non-topological insulator phases.
We also discuss the analogy between the phase structure found here and that of lattice quantum chromodynamics (QCD).
It is known that lattice QCD with a certain lattice fermion formalism possesses a {parity-broken} phase
similar to the tilted AF phase in the strong coupling region,
which is called ``Aoki phase'' \cite{Aoki}.
From the analogy between these phases,
we can give a conjecture on the phase structure of topological insulators to some extent,
from the well-known phase structure of lattice QCD.

This paper is organized as follows.
In Section \ref{sec:noninteracting},
we review the band theory and the topological phase structure of graphene and topological insulators (Kane--Mel\'e) model,
in the absence of electron-electron interaction.
In Section \ref{sec:lattice},
we construct an effective U(1) gauge theory on the honeycomb lattice,
to incorporate the electron-electron interaction in terms of QED.
In Section \ref{sec:strong-coupling},
we apply the techniques of strong coupling expansion to the gauge theory on the honeycomb lattice,
and the behavior of the AF order is investigated in the strong coupling limit of the interaction.
As a result, we obtain the topological phase diagram under the electron-electron interaction,
with a new ``tilted AF'' phase.
In Section \ref{sec:discussion},
we discuss the physical properties of the system in the tilted AF phase.
We also compare this phase structure to that of lattice QCD,
and summarize the analogy between them.
Finally, in Section \ref{sec:summary},
we conclude our study and raise several open questions.

\section{Band theory of noninteracting systems}\label{sec:noninteracting}

\begin{figure}[tbp]
\begin{center}
\begin{tabular}{ccc}
\includegraphics[width=3cm]{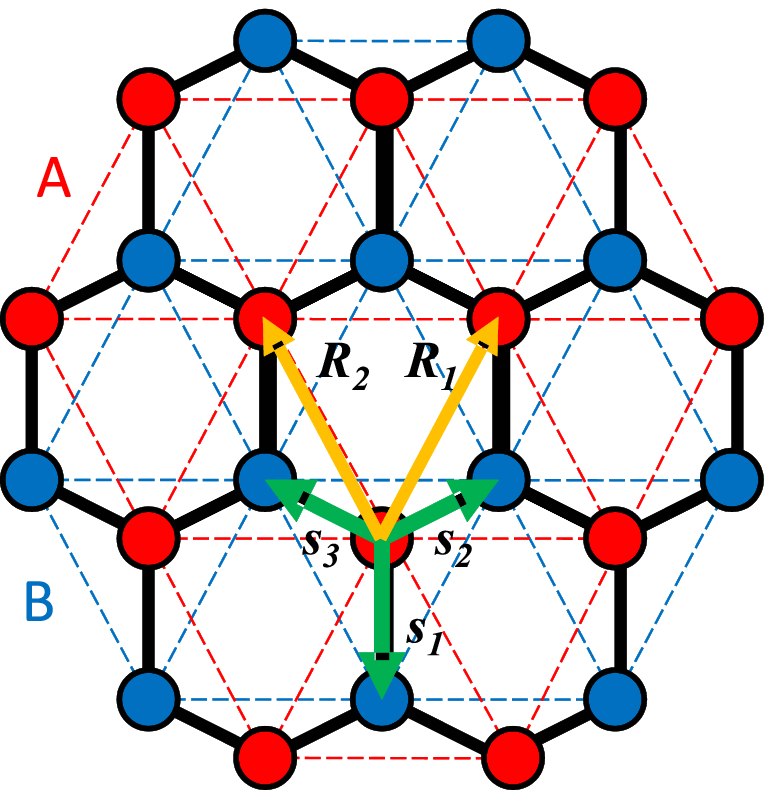}
 & 
\includegraphics[width=2.5cm]{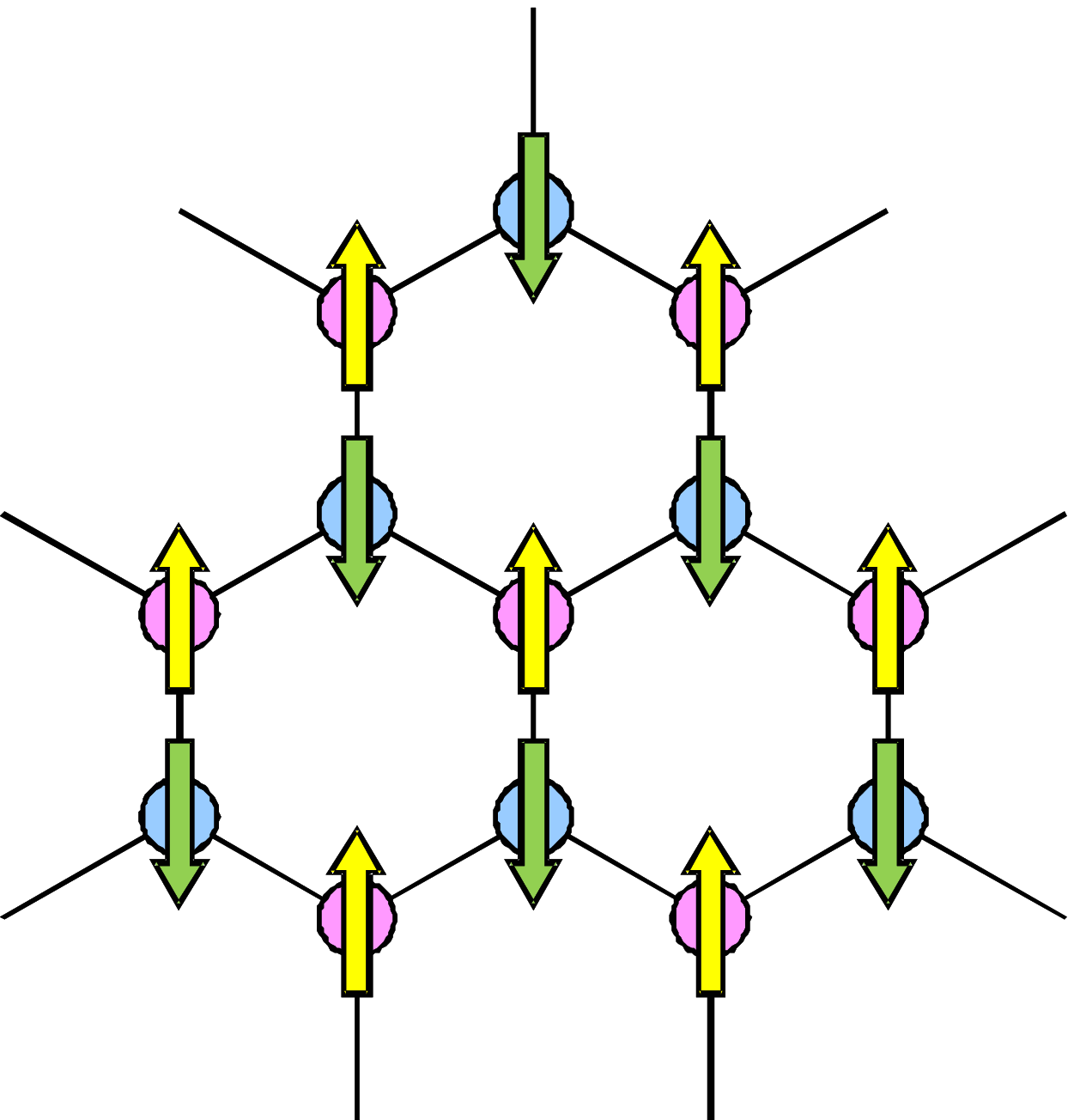}
 &
\includegraphics[width=2.5cm]{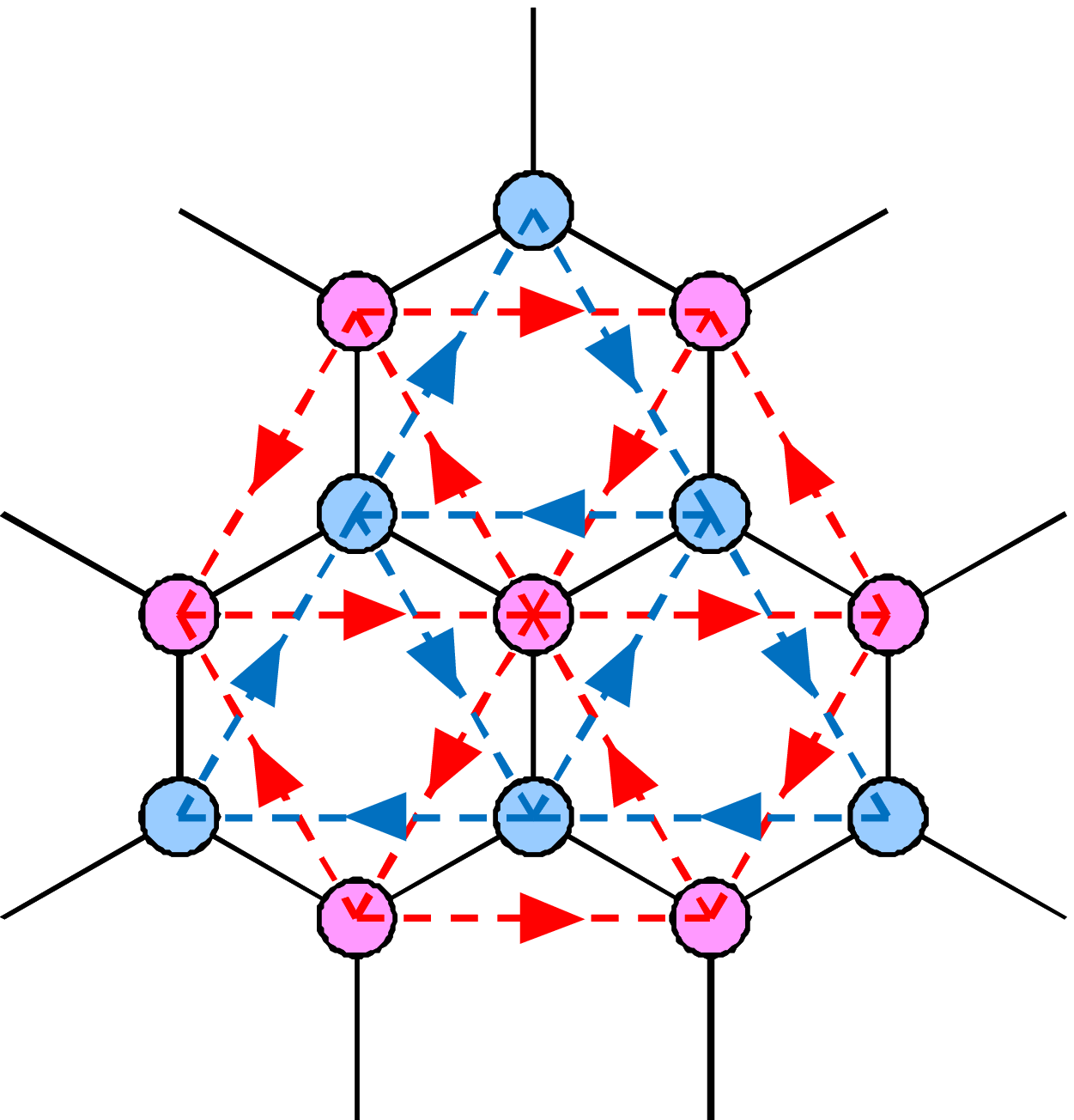}
 \\
(a) & (b) & (c) \\
\end{tabular}
\end{center}
\caption{(a) Configuration of the honeycomb lattice and its sublattice structure. (b) Schematic picture of the staggered magnetic field. (c) Schematic picture of the Kane--Mel\'e model. The spin-orbit interaction is introduced in terms of a complex hopping between next-to nearest-neighboring sites, depending on its direction and spin.}
\label{fig:configuration}
\end{figure}

Before starting the discussion on the electron-electron interaction effects,
let us briefly review the band theory and the topological phase structure of
graphenelike systems, including topological insulators, without any interaction.
The honeycomb lattice consists of two triangular sublattices, A and B,
both of which are spanned by lattice vectors $\bfR_1=\bfs_2-\bfs_1$ and $\bfR_2=\bfs_3-\bfs_1$,
with $\bfs_{i=1,2,3}$ the vectors connecting every nearest-neighboring (NN) sites (see Fig.\ref{fig:configuration}(a)).
The dynamics of electrons on the lattice is described by the conventional tight-binding Hamiltonian,
$H_T = -t\sum_{\bfr_A,\sigma,i}\left[a_\sigma^\dag(\bfr_A) b_\sigma(\bfr_A+\bfs_i)+\mathrm{H.c.}\right]$,
where $a(\bfr_A),b(\bfr_B)$ are annihilation operators for A and B sites respectively,
and the sum by $\sigma=\uparrow,\downarrow$ assures the SU(2) spin symmetry.
This Hamiltonian reads $H_T = -t \sum_{\bfk \in \Omega} \left[b^\dag(\bfk) \Phi(\bfk) a(\bfk) +\mathrm{H.c.}\right]$
in the momentum space,
so that the eigenvalue is given as $E(\bfk)=\pm t|\Phi(\bfk)|$,
where the momentum kernel $\Phi$ reads $\Phi(\bfk) =\sum_{i} e^{i\bfk\cdot\bfs_i}$.
This band structure reveals the well-known Dirac cone (valley) structure around two Dirac points $\bfK_\pm$
in the Brillouin zone $\Omega$ \cite{wallace_1947}.

One way to open a finite bandgap at the Dirac points is an application of a ``staggered magnetic field'',
$ H_M = m \sum \left[a^\dag \sigma_z a - b^\dag \sigma_z b \right]$,
which favors the down spin component at A sites while up at B sites (see Fig.\ref{fig:configuration}(b)).
This term explicitly breaks the SU(2) spin symmetry and the sublattice (exchange) symmetry,
serving as a mass term for the (4-component) Dirac fermion.
It opens a bandgap $|m|$ at both Dirac points,
keeping the topology of the ground state wave function trivial.

On the other hand, the spin-orbit interaction opens a finite gap
accompanied with a topologically nontrivial ground state.
Following Kane and Mel\'e \cite{Kane-Mele},
it is given 
in terms of a complex hopping term between next-to NN (NNN) sites on the honeycomb lattice (see Fig.\ref{fig:configuration}(c)),
\begin{align}
H_{SO} = t' \!\!\!\!\!\! \sum_{\langle \langle \bfr_A,\bfr'_A \rangle \rangle} \!\!\!\!\!\! e^{\pm i\phi} a^\dag(\bfr_A) \sigma_z a(\bfr'_A) 
 + t' \!\!\!\!\!\! \sum_{\langle \langle \bfr_B,\bfr'_B \rangle \rangle} \!\!\!\!\!\! e^{\pm i\phi} b^\dag(\bfr_B) \sigma_z b(\bfr'_B),
\end{align}
where the sum is taken over all the pairs of NNN sites
$\langle \langle \bfr_A,\bfr'_A \rangle \rangle$ or $\langle \langle \bfr_B,\bfr'_B \rangle \rangle$.
The phase ${\pm i\phi}$ takes the plus sign in the direction pointed by the arrows in Fig.\ref{fig:configuration}(c),
while the minus sign in the opposite direction.
If we fix the phase $\phi=\pi/2$,
the Hamiltonian reads
\begin{align}
H_{SO} = - \sum_{\bfk \in \Omega} 2t' \im \Phi_2(\bfk) \left[a^\dag(\bfk)\sigma_z a(\bfk) -b^\dag(\bfk)\sigma_z b(\bfk)\right]
\end{align}
in the momentum space,
where $\Phi_2$ is defined by $\Phi_2(\bfk) = e^{i\bfk\cdot\bfR_1} + e^{-i\bfk\cdot\bfR_2} + e^{i\bfk\cdot(\bfR_2-\bfR_1)}$.
Similar to the staggered magnetic field,
it breaks the sublattice and spin symmetry,
while it also breaks the exchange symmetry of two valleys in addition.
Thus, it opens a bandgap with an amplitude $|3\sqrt{3}t'|$ for each valley
as an ``effective mass'' for Dirac fermion,
with its sign depending on the valley/spin indices.
As a result, the ground state wave function acquires a nontrivial topology
with a non-zero $Z_2$ topological number given in the momentum space,
leading to a quantized spin Hall conductivity.

The topological phase structure of the system is characterized by
the competition between the conventional gap and the topological gap.
Applying both gap opening effects given above,
one valley obtains a bandgap with amplitude $m+3\sqrt{3}t'$,
while the other $m-3\sqrt{3}t'$.
When the conventional gap is dominant over the topological one ($|m|>|3\sqrt{3}t'|$),
both valleys obtain a bandgap with the same sign, leaving the ground state trivial.
When the spin-orbit interaction is dominant ($|m|<|3\sqrt{3}t'|$), 
the system behaves as a topological insulator.
Two phases are separated by a line $m=\pm 3\sqrt{3}t'$,
on which one of the valleys loses its gap while the other remains gapped.
In this paper, we focus on how the electron-electron interaction alters such a topological phase structure.

\section{Lattice gauge theory description}\label{sec:lattice}

Here we assume that the electron-electron interaction is mediated by the electromagnetic field,
namely U(1) gauge field $A_{\mu=0,1,2,3}$.
The dynamics of fermions can be reconstructed in terms of an {imaginary time action} on the honeycomb lattice
(see Appendix \ref{app:path-integral} for details),
\begin{align}
S_F=&\frac{1}{2}\sum_{\bfr_A,\tau}\left[a^\dag(\bfr_A,\tau)U_0(\bfr_A,\tau)a(\bfr_A,\tau+\Delta\tau) -\mathrm{H.c.}\right] \nonumber \\
 & +\frac{1}{2}\sum_{\bfr_B,\tau}\left[b^\dag(\bfr_B,\tau)U_0(\bfr_B,\tau)b(\bfr_B,\tau+\Delta\tau) -\mathrm{H.c.}\right] \nonumber \\
& +\frac{t\Delta\tau}{\vf}\sum_{\bfr_A,i,\tau}\left[a^\dag(\bfr_A,\tau) b(\bfr_A+\bfs_i,\tau)+\mathrm{H.c.}\right], \label{eq:s-f}
\end{align}
with the U(1) link variables $U_0(\bfr,\tau)=\exp\left[ie\int_{\tau}^{\tau+\Delta\tau} d\tau' A_0(\bfr,\tau')\right]$.
Dynamics of the spatial components of electromagnetic field, $U_i$ (or $A_i$), is neglected,
which is referred to as ``instantaneous approximation'' \cite{Son_2007},
since the gauge field propagates much faster than the fermions,
so that the retardation effect becomes considerably small.
Here the imaginary time direction is rescaled by the Fermi velocity $\vf$ to avoid the space-time anisotropy in the Dirac operator,
and is discretized by the lattice spacing $\Delta\tau$ comparable to the spatial lattice constant $a=|\bfs_i|$.
Since this discretization leads to a pair of fermion doublers in the temporal direction \cite{Nielsen-Ninomiya},
here we suppress the spin index $\sigma$ to attribute the doublers to the realistic spin degrees of freedom,
like in the staggered fermion formalism \cite{Susskind_1977,Sharatchandra_1981}.
The lattice action Eq.(\ref{eq:s-f}) is invariant under the global U(1) charge transformation
$a \rightarrow e^{i\theta} a, \quad a^\dag \rightarrow a^\dag e^{-i\theta},
\quad b \rightarrow e^{i\theta} b, \quad b^\dag \rightarrow b^\dag e^{-i\theta}$,
and the U(1) spin transformation
\begin{align}
a_\mathrm{e} \rightarrow e^{i\tilde{\theta}} a_\mathrm{e},\; 
a^\dag_\mathrm{e} \rightarrow a^\dag_\mathrm{e}e^{i\tilde{\theta}},\; &
b_\mathrm{e} \rightarrow e^{-i\tilde{\theta}}b_\mathrm{e},\;
b^\dag_\mathrm{e} \rightarrow b^\dag_\mathrm{e}e^{-i\tilde{\theta}}, \nonumber \\
a_\mathrm{o} \rightarrow e^{-i\tilde{\theta}}a_\mathrm{o},\; 
a^\dag_\mathrm{o} \rightarrow a^\dag_\mathrm{o}e^{-i\tilde{\theta}},\; &
b_\mathrm{o} \rightarrow e^{i\tilde{\theta}}b_\mathrm{o},\;
b^\dag_\mathrm{o} \rightarrow b^\dag_\mathrm{o}e^{i\tilde{\theta}}, \label{eq:remnant}
\end{align}
where the label $\mathrm{e/o}$ represents whether the {discretized} imaginary time $\tau/\Delta\tau$ is even or odd.
It should be noted that spin SU(2) symmetry is broken down to U(1) subspace due to the temporal discretization,
which is analogous to the intrinsic flavor (taste) symmetry breaking in the staggered fermion formulation.
{Similarly to the staggered fermion,
the full spin symmetry is restored in the continuum limit $(\tau/\Delta\tau \rightarrow 0)$.}
Since we can choose the spin direction arbitrarily,
here we define the remnant U(1) spin symmetry in the $(x,z)$-plane,
{generated by the spin operator $\sigma_y$}.

The spin-orbit interaction term $H_{SO}$ and the staggered magnetic field $H_M$
are reconstructed in the path integral formalism as
\begin{align}
S_{SO} =& t'\frac{\Delta\tau}{\vf} \Biggl[ \sum_{\langle \langle \bfr_A,\bfr'_A \rangle \rangle ,\tau} \pm i a^\dag(\bfr_A,\tau) a(\bfr'_A,\tau) \nonumber \\
 & \quad\quad + \sum_{\langle \langle \bfr_B,\bfr'_B \rangle \rangle,\tau} \pm i b^\dag(\bfr_B,\tau) b(\bfr'_B,\tau) \Biggr], \\
S_{M} =& m\frac{\Delta\tau}{\vf} \left[\sum_{\bfr_A,\tau} a^\dag a - \sum_{\bfr_B,\tau} b^\dag b\right],
\end{align}
where both terms break the sublattice symmetry and the remnant U(1) spin symmetry explicitly.
Hereafter we suppress the rescaling factor $\Delta\tau/\vf$ for simplicity,
so that we regard the parameters $t$, $t'$ and $m$ dimensionless.

Dynamics of the gauge field can also be defined on the honeycomb lattice,
in terms of the polynomial of link variables $U_0$.
The gauge kinetic term $S_G$ is proportional to the parameter $\beta=\epsilon_0 \vf/e^2$,
which corresponds to the inverse of the (effective) Coulomb coupling strength.
When the Fermi velocity of the electron $\vf$ is sufficiently small compared to the speed of light,
the effective coupling $\alpha=e^2/4\pi\epsilon_0\vf$ becomes larger than the usual $\alpha_\mathrm{QED}=e^2/4\pi\epsilon_0 c\sim 1/137$,
since a slower electron feels the effect of the electromagnetic field more strongly.
The parameter $\beta$ becomes quite small in such a system.
For instance, $\beta \sim 0.04$ in vacuum-suspended graphene,
where $\vf$ is about 300 times smaller than the speed of light.
$S_G$ vanishes in the strong coupling limit $\beta=0$,
i.e. the spatial propagation of the electromagnetic field is completely suppressed.

\section{Strong coupling analysis}\label{sec:strong-coupling}
\subsection{Antiferromagnetism in the non-topological system}
Let us first review the behavior of the non-topological system,
in the absence of the spin-orbit interaction and the staggered magnetic field,
in the strong coupling limit $\beta=0$.
In this limit,
we can rewrite this effective action only in terms of fermionic field variables
by integrating out the gauge degrees of freedom,
\begin{align}
& S_F^{(0)} = t \sum_{\bfr_A,i,\tau}\left[a^\dag(\bfr_A,\tau) b(\bfr_A+\bfs_i,\tau)+\mathrm{H.c.}\right] \\
& -\frac{1}{4}\left[\sum_{\bfr_A,\tau}n_{\bfr_A}(\tau)n_{\bfr_A}(\tau+\Delta\tau) + \sum_{\bfr_B,\tau}n_{\bfr_B}(\tau)n_{\bfr_B}(\tau+\Delta\tau)\right], \nonumber
\end{align}
where $n_{\bfr_A}(\tau)=a^\dag(\bfr_A,\tau)a(\bfr_A,\tau)$ and $n_{\bfr_B}(\tau)=b^\dag(\bfr_B,\tau)b(\bfr_B,\tau)$
denote the local charge density at time $\tau$.
In the leading order of the strong coupling expansion,
an on-site interaction with a temporal lattice spacing is extracted from the long-range Coulomb interaction,
which is similar to the on-site repulsion term in the phenomenological Hubbard model.

Respecting the sublattice symmetry and the remnant U(1) spin symmetry,
here we take the mean-field ansatz
\begin{align}
\langle a^\dag(\bfr_A,\tau) a(\bfr_A,\tau)\rangle &= \tfrac{1}{2}\left[\sigma_1 -i(-1)^{\tau/\Delta\tau}\sigma_2\right] \\
\langle b^\dag(\bfr_B,\tau) b(\bfr_B,\tau)\rangle &= \tfrac{1}{2}\left[-\sigma_1 -i(-1)^{\tau/\Delta\tau}\sigma_2\right],
\end{align}
where $\sigma_{1,2}$ are real values.
{(One should not confuse the mean fields $\sigma_{1,2}$ with the Pauli matrices $\sigma_{x,y,z}$.)}
Thus, by integrating out the fermionic field variables,
we obtain the thermodynamic potential (free energy) of the system per a pair of A and B sites,
\begin{align}
F_\mathrm{eff}(\sigma) = \frac{1}{2}|\sigma|^2 - \int_{\Omega} d^2\bfk \ln\left[\frac{|\sigma|^2}{4} +|t\Phi(\bfk)|^2\right], \label{eq:f-0}
\end{align}
where the momentum integration within the Brillouin zone $\Omega$ is normalized as $\int_\Omega d^2\bfk =1$.
Here the order parameter $\sigma=\sigma_1+i\sigma_2$ appears only in the form of $|\sigma|^2=\sigma_1^2+\sigma_2^2$,
which reflects the remnant U(1) spin symmetry,
$ \sigma \rightarrow \sigma e^{2i\tilde{\theta}}$.
This symmetry gets broken spontaneously when $|\sigma|$ takes a finite expectation value,
but the phase of $\sigma$ can be chosen arbitrarily
unless the symmetry-breaking source $S_{SO}$ or $S_M$
is applied.
The arbitrariness results in the emergence of gapless Nambu--Goldstone boson
when the $\mathrm{U(1)_V}$ symmetry is spontaneously broken.
$\sigma$ serves as the order parameter for the spontaneous breaking of the sublattice symmetry and the remnant U(1) spin symmetry,
which corresponds to the spin density wave (SDW) order,
or the antiferromagnetism, on the honeycomb lattice.

The first term in Eq.(\ref{eq:f-0}),
which comes from the tree level of the bosonic auxiliary field $\sigma$,
becomes dominant for $|\sigma| \rightarrow \infty$,
while the second term,
which stems from the fermion one-loop effect,
yields a logarithmic singularity around $\sigma=0$.
Therefore, $F_\mathrm{eff}(\sigma)$ possesses a minimum at finite $|\sigma|$,
so that there appears a spontaneous antiferromagnetic order with an arbitrary direction in the remnant U(1) spin space.
This antiferromagnetism opens a finite bandgap
in terms of a dynamical ``mass term'',
while the $Z_2$ topology of the system remains unchanged.

\subsection{Effect of interaction in the topological system}
Let us now investigate how the spontaneous antiferromagnetism obtained above behaves
in the presence of the (Kane--Mel\'e-type) spin-orbit interaction $S_{SO}$.
We also introduce the uniform staggered magnetic field $S_M$ for convenience of later analysis.
Since the spin-orbit interaction and the staggered magnetic field explicitly break the remnant U(1) spin symmetry
in the $\sigma_1$-direction,
the order parameter components $\sigma_1$ and $\sigma_2$ should be distinguished here.
Here the effective potential in the strong coupling limit reads
\begin{align}
F_\mathrm{eff}(\sigma) = \frac{(\sigma_1-2m)^2+\sigma_2^2}{2} -\int_{\Omega} d^2\bfk \ln \left[\mathcal{E}(\sigma_1,\sigma_2,t';\bfk)\right]^2
\end{align}
where we have shifted the order parameter $\sigma_1/2+m \rightarrow \sigma_1/2$,
which serves as the ``modified'' source $m$ due to the electron-electron interaction.
$ \mathcal{E}(\sigma_1,\sigma_2,t';\bfk) = \sqrt{\left[\sigma_1/2-2t' \im\Phi_2(\bfk) \right]^2 + \left(\sigma_2/2 \right)^2+|t\Phi(\bfk)|^2} $ denotes the energy of an electron in the conduction band.
The minimum of this effective potential $(\tilde{\sigma}_1,\tilde{\sigma}_2)$ satisfies the gap equations
\begin{align}
\frac{\partial F_\mathrm{eff}}{\partial \sigma_1} \Bigr|_{(\tilde{\sigma}_1,\tilde{\sigma}_2)} &= \tilde{\sigma}_1 -2m -\int_{\Omega} d^2\bfk \frac{\tilde{\sigma}_1/2-2t'\im\Phi_2(\bfk)}{\left[\mathcal{E}(\tilde{\sigma}_1,\tilde{\sigma}_2,t';\bfk)\right]^2} =0 \label{eq:gapeq-1}\\
\frac{\partial F_\mathrm{eff}}{\partial \sigma_2} \Bigr|_{(\tilde{\sigma}_1,\tilde{\sigma}_2)} &= \tilde{\sigma}_2 -\int_{\Omega} d^2\bfk \frac{\tilde{\sigma}_2/2}{\left[\mathcal{E}(\tilde{\sigma}_1,\tilde{\sigma}_2,t';\bfk)\right]^2} =0. \label{eq:gapeq-2}
\end{align}
Since the potential is even in $\sigma_2$,
the solution should satisfy either $\tilde{\sigma}_2=0$,
where the antiferromagnetic order is aligned in the $\sigma_1$-direction,
 or $(1/\tilde{\sigma}_2)(\partial F_\mathrm{eff}/\partial \sigma_2)|_{(\tilde{\sigma}_1,\tilde{\sigma}_2)}=0$, where it is tilted toward the $\sigma_2$-direction.

\begin{figure}[tbp]
\begin{center}
\includegraphics[width=7cm]{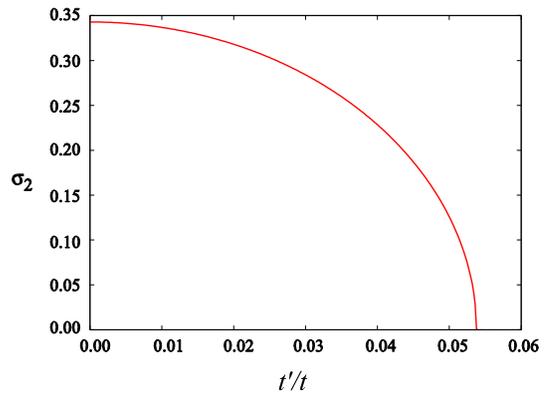}
\end{center}
\caption{The behavior of the order parameter $\sigma_2$ as a function of the amplitude $t'$ of spin-orbit interaction,
in the absence of explicit mass $m$.
$\sigma_2$ vanishes at the critical value $t'_C=0.0538t$.}
\label{fig:haldane_wilson_massless}
\end{figure}

First, we consider the case in the absence of the staggered potential $m$,
where the effective potential becomes even both in $\sigma_1$ and $\sigma_2$.
In this case the potential minimum satisfies $\sigma_1=0$,
so that the antiferromagnetic order points the $\sigma_2$-direction,
i.e. it is confined in the $xy$-plane (see Appendix \ref{app:sigma=0} for detail).
Here it should be noted that,
even though the spin-orbit interaction explicitly breaks the remnant U(1) spin symmetry in the $\sigma_1$-direction,
the antiferromagnetic order is aligned in the $\sigma_2$-direction, orthogonally to $\sigma_1$.

The quantitative behavior of $\sigma_2$ is obtained by minimizing the effective potential
\begin{align}
F_\mathrm{eff}(\sigma_2) = \frac{\sigma_2^2}{2} -\int_{\Omega} d^2\bfk \ln \left[\mathcal{E}(0,\sigma_2,t';\bfk)\right]^2,
\end{align}
by $\sigma_2$.
Since the finite bandgap $3\sqrt{3}t'$ from the spin-orbit interaction at each Dirac point
moderates the logarithmic singularity in the loop integral,
it suppresses the expectation value of $\sigma_2$.
Second order phase transition occurs at the critical value $t'_C=0.0538t$,
as shown in Fig.\ref{fig:haldane_wilson_massless}.

\begin{figure}[tbp]
\begin{center}
\includegraphics[width=8cm]{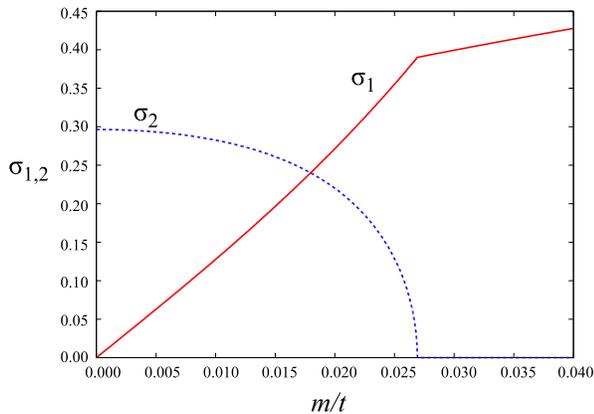}
\end{center}
\caption{Behavior of the order parameters $\sigma_{1,2}$ as a function of $m$,
with the amplitude of spin-orbit interaction fixed at $t'=0.5t'_C$.
The staggered magnetic field $m$ monotonically enhances $\sigma_1$ as its external source,
while it suppresses the orthogonal order parameter $\sigma_2$.}
\label{fig:haldane-s12}
\end{figure}

\begin{figure}[tbp]
\begin{center}
\includegraphics[width=8cm]{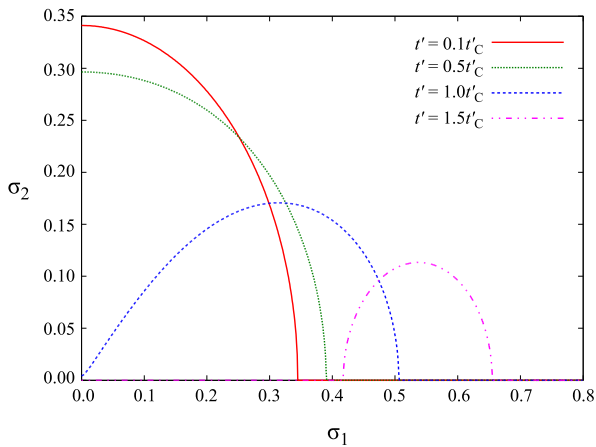}
\end{center}
\caption{The path of order parameter $(\sigma_1,\sigma_2)$,
with the spin-orbit interaction $t'$ fixed and the staggered magnetic field $m$ varied.
The path starts from $\sigma_1=0$ at $m=0$,
and evolves to $\sigma_1 \rightarrow \infty$ monotonically as $m \rightarrow \infty$,
since the staggered magnetic field $m$ serves as the external source for the antiferromagnetic order
in the $\sigma_1$-direction.
}
\label{fig:haldane-path}
\end{figure}

Next, we fix the spin-orbit interaction $t'$ and introduce the uniform staggered magnetic field $S_M$.
For instance, by fixing $t'=0.5t'_C$,
we can observe the behavior of the order parameters $\sigma_{1,2}$ as a function of $m$,
as shown in Fig.\ref{fig:haldane-s12}.
Since $S_M$ serves as a source term for the AF order in the $\sigma_1$-direction,
it eventually tilts the direction of $\sigma$ from the $\sigma_2$-axis
toward the $\sigma_1$-axis.
Path of the solution $(\sigma_1,\sigma_2)$, with $t'$ fixed and $m$ varied, is displayed in Fig.\ref{fig:haldane-path}
(see Appendix \ref{app:evolution} for details).
Starting from $\sigma_1=0$ at $m=0$,
$\sigma_1$ monotonically increases as a function of $m$.
For $t'<t'_C$,
the AF order gets tilted from $\sigma_2$-axis to the $\sigma_1$-axis,
and finally $\sigma_2$ vanishes at some critical value of $m$ depending on $t'$.
On the other hand, if $t'>t'_C$, the path starts from $\sigma_1=\sigma_2=0$,
and $\sigma_1$ evolves with the staggered magnetic field $m$.
In both cases, when $m$ reaches a sufficiently large value,
the electron-electron interaction can be neglected compared to the explicit gap $m$,
so that the AF order $\sigma$ is aligned in the $\sigma_1$-direction,
parallel to the spin-orbit interaction and the staggered magnetic field.

\begin{figure}[tbp]
\begin{center}
\includegraphics[width=8cm]{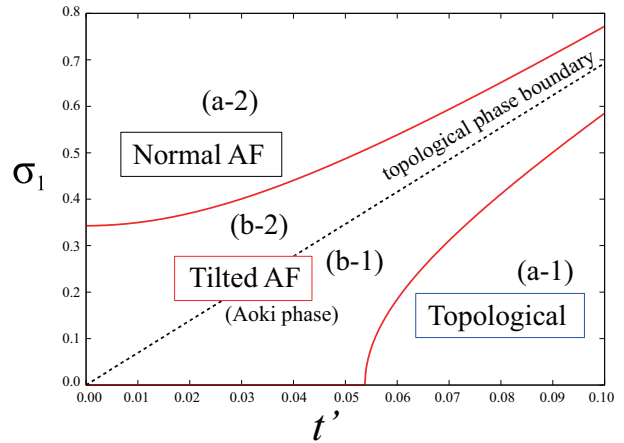}
\end{center}
\caption{The phase diagram of this system in the $(t',\sigma_1)$-space.
Here we use the ``modified'' mass $\sigma_1$ instead of the bare mass (staggered magnetic field) $m$,
since there is a unique one-to-one correspondence between $m$ and $\sigma_1$, depending on $t'$.
As a consequence of the interplay between the electron-electron interaction and the spin-orbit interaction,
there appears a new ``tilted antiferromagnetic (AF)'' phase ($\sigma_2 \neq 0$),
between the normal AF phase ($\sigma_2=0$ and $\sigma_1/2 > 3\sqrt{3}t'$)
and the topological phase ($\sigma_2=0$ and $\sigma_1/2 > 3\sqrt{3}t'$).
}
\label{fig:haldane-phasediagram}
\end{figure}

\begin{figure}[tbp]
\begin{center}
\includegraphics[width=8cm]{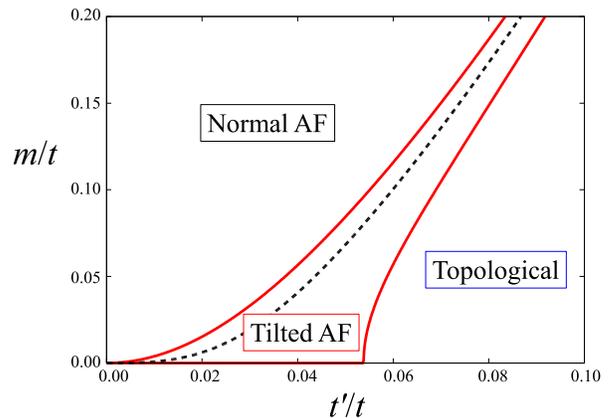}
\end{center}
\caption{The phase diagram in Fig.\ref{fig:haldane-phasediagram} mapped in the $(t',m)$-plane.
The dashed line corresponds to the topological phase boundary $\tilde{\sigma}_1/2 = 3\sqrt{3}t'$.
The tilted AF phase converges to a point at $(t',m)=(0,0)$.}
\label{fig:haldane-phasediagram-new}
\end{figure}

Thus we can map the phase diagram of this system in the parameter space $(t',m)$.
Since $\sigma_1$ monotonically increases as a function of $m$ (depending on $t'$),
here we take $\sigma_1$ as a control parameter instead of $m$.
There are two branches of phase boundary between the phase with finite $\sigma_2$ and that with $\sigma_2=0$:
one is present for the whole range of $t'$,
while the other is restricted in the region $t'\geq t'_C$.
When the explicit bandgap from $t'$ and $m$ is extremely large
compared to the scale of the electron-electron interaction,
the logarithmic singularity in the fermion loop integral becomes dominant
only around the the topological phase boundary $\sigma_1/2 = 3\sqrt{3}t'$,
where one of the Dirac cones loses its bandgap.
Therefore, the $\sigma_2\neq 0$ phase shrinks to the topological phase boundary in this limit,
so that the phase boundaries characterized by $\sigma_2$ discussed above
approach asymptotically along the topological phase boundary.
As a consequence, the phase structure of this system is classified into three phases, as shown in Fig.\ref{fig:haldane-phasediagram}:
\begin{itemize}
\item ``Topological'' phase ($\sigma_1/2 < 3\sqrt{3}t'$ and $\sigma_2=0$):
The AF order, aligned in the $\sigma_1$-direction,
is rather small compared to the explicit gap given by the spin-orbit interaction.
Thus, the system becomes a $Z_2$ topological insulator,
yielding the quantum spin Hall effect even under the electron-electron interaction.
\item ``Normal AF'' phase ($\sigma_1/2 > 3\sqrt{3}t'$ and $\sigma_2=0$):
The commensurate AF order $\sigma_1$
exceeds the explicit gap given by the spin-orbit interaction $t'$,
so that the system becomes a conventional (non-topological) insulator.
\item ``Tilted AF'' phase ($\sigma_2 \neq 0$):
The AF order is tilted from $\sigma_1$-axis toward the $\sigma_2$-axis.
\end{itemize}
One should be careful of the tilted AF region at $t'=0$.
This region can be reached at $t'=m=0$,
where the remnant U(1) spin symmetry is not explicitly broken.
Since $(\sigma_1,\sigma_2)$ can be chosen arbitrarily with keeping $|\sigma|^2=\sigma_1^2+\sigma_2^2$ constant,
here the ground state can take any point within this region.
If we map it by the original parameter set $(t',m)$, as shown in Fig.\ref{fig:haldane-phasediagram-new},
such a region corresponds to the original point of the phase diagram.

As a result, in the presence of the electron-electron interaction,
the phase structure of the system is altered from the noninteracting system,
with the emergence of ``tilted AF'' phase between the topological phase and the conventional insulator phase.
In other words,
the topological phase ``boundary'' in the noninteracting system evolves into the tilted AF ``region''
by the effect of the electron-electron interaction.
We shall discuss the physical properties of this phase in the next section.

\section{Discussion}\label{sec:discussion}

\subsection{Physical properties of the tilted AF phase}
\begin{figure}[tbp]
\begin{center}
\includegraphics[width=6cm]{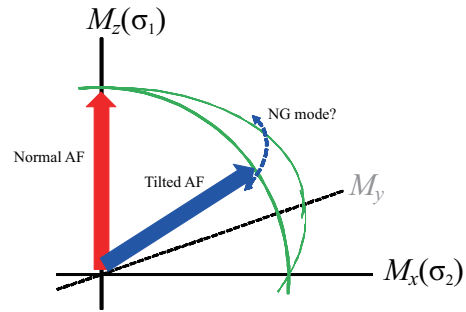}
\end{center}
\caption{Schematic picture of the order parameters derived in this study.
$\sigma_1$ and $\sigma_2$ are antiferromagnetic (AF) orders corresponding to two directions in the remnant U(1) spin space,
which we denote $M_z$ and $M_x$ here.
The staggered magnetic field $m$ explicitly breaks this symmetry to the $\sigma_1$-direction.
When $\langle \sigma_2 \rangle \neq 0$,
the AF order is tilted to the $\sigma_2$-direction to some extent.
If we extend this argument to the full SU(2) spin space,
another direction $M_y$ is restored,
so that the tilted AF acquires U(1) degree of freedom in choosing its direction,
which may result in a massless Nambu--Goldstone mode.}
\label{fig:modes}
\end{figure}

What does the emergence of the ``tilted AF'' phase physically imply?
We should recall that the order parameters $\sigma_1$ and $\sigma_2$ correspond to
the AF order in the $z$- and $x$- (or $y$-)directions respectively,
breaking the remnant U(1) spin symmetry.
In the tilted AF phase,
the direction of the AF order is tilted from $z$-axis, pointed by the spin-orbit interaction and the staggered magnetic field,
to the $xy$-plane,
by the interplay between the electron-electron interaction and the spin-orbit interaction,
as sketched in Fig.\ref{fig:modes}.
In the absence of the staggered magnetic field $m$,
$\sigma$ is completely tilted to the $\sigma_2$-direction,
which is consistent with the ``XY-antiferromagnetic insulator'' phase
found in the analysis of the Kane--Mel\'e--Hubbard model \cite{Muramatsu_Assaad,Reuther,Vaezi}.

If we restore the spin space from remnant U(1) to full SU(2),
there appears a U(1) degree of freedom in choosing the direction of the AF order within the $xy$-plane,
which results in the appearance of a gapless Nambu--Goldstone mode,
while the fluctuation in the $z$-direction becomes massive.
Thus, although the fermion spectrum is gapped,
it is possible that such a gapless boson may carry an electric current,
turning the system back from insulator into a (semi-)metal.
On the other hand, when the antiferromagnetic order is aligned in the $z$-direction (``normal AF'' phase),
the phase fluctuation of the order parameter will result in two massive modes,
so that the system remains an insulator.
Phase transition between the topological insulator (QSH phase) and the metallic phase
has also been suggested in quantum Hall systems, driven by a disorder \cite{Beenakker}.

In the phase diagram obtained here,
the phase transition between the topological phase and the conventional insulating phase
occurs without closing the bandgap of fermions.
This behavior appears to contradict the previous studies in the noninteracting Dirac fermion system,
where the gap closing is essential for the topological phase transition \cite{Murakami_2007,Murakami_2011}.
In this study, however,
one should note the appearance of a gapless NG mode in the tilted AF phase,
which is not taken into account in the noninteracting system.
The qualitative properties and physical effects of this NG mode,
which cannot be investigated within our analysis due to the restriction of the spin space as a lattice artifact,
remains a future problem.

\begin{table*}
\begin{center}
\begin{tabular}{|c|c|c|}
\hline
 & \textbf{Kane--Mel\'e model} & \textbf{Lattice QCD} \\
\hline \hline
Non-topological source & staggered magnetic field $(m)$ & mass term \\
\hline
Topological source & spin-orbit interaction $(t')$ & Wilson term \\
\hline
Splits the degeneracy of & valleys & doublers \\
\hline
Explicitly breaks & remnant U(1) spin symmetry & continuous chiral symmetry \\
\hline
Interaction is mediated by & photons (electromagnetic field) & gluons \\
\hline
Induced phase & tilted AF phase & Aoki phase \\
\hline
Order parameter & $\langle \sigma_2 \rangle (\sim \langle a^\dag \sigma_x a - b^\dag \sigma_x b\rangle)$ & $\langle \bar{\psi} i\gamma_5 \psi \rangle$ (pion condensation) \\
\hline
\end{tabular}
\caption{Analogy between the Kane--Mel\'e model (2-dimensional topological insulators) and lattice QCD with Wilson fermion.}
\label{table:wilson}
\end{center}
\end{table*}

\subsection{Analogy with the phase structure of lattice QCD}
The phase structure of 2-dimensional topological insulators observed here can be understood
in analogy with that of lattice QCD, which has been thoroughly studied by strong coupling expansion,
Monte Carlo simulations, etc.
In order to avoid the doubling of quarks arising from the lattice discretization,
one can take the Wilson fermion formalism,
where a momentum-dependent mass term (Wilson term) is employed in addition to the uniform mass term,
to shift the degeneracy of doublers \cite{Wilson-fermion}.
One can extract a single species of fermion with the lowest effective mass
out of the doublers,
by taking the Wilson parameter sufficiently large.
This mechanism is analogous to the spin-orbit interaction on the honeycomb lattice,
which shifts the degeneracy of two valleys in the Brillouin zone $\Omega$.
Of course, the continuous chiral symmetry of the quarks,
which corresponds to the remnant U(1) spin symmetry in our graphene model,
is explicitly broken by this effective mass term.

It is known that lattice QCD with the Wilson fermion formalism has a characteristic phase structure.
In the strongly coupled QCD,
the chiral symmetry is spontaneously broken with a finite chiral condensate $\langle \bar{\psi} \psi \rangle$,
dynamically generating a mass of quarks.
The mass term $m \bar{\psi}\psi$ serves as a source term for the chiral condensate.
In lattice QCD with a {single-flavor} Wilson fermion, on the other hand,
there appears a finite pion condensate $\langle \bar{\psi} i\gamma_5 \psi \rangle$,
which is orthogonal to the chiral condensate $\langle \bar{\psi} \psi \rangle$
 in the chiral symmetry space \cite{Aoki}.
This phase is called ``Aoki phase'',
where the {parity symmetry} is spontaneously broken by the pion condensation.
{In the two-flavor theory,
the Aoki phase is characterized by a neutral pion condensate
$\langle \bar{\psi} i(\gamma_5 \otimes \tau_z) \psi \rangle$,
where the parity-flavor symmetry is broken instead of the parity symmetry itself
($\tau_z$ is a Pauli matrix with respect to the ``isospin'', corresponding to the flavor of quarks).}
If we fix the Wilson parameter and give a sufficiently large mass term uniformly to all the doublers,
the pion condensate disappears, and only the chiral condensate acquires a finite expectation value.
It is known that the transition between the Aoki phase and the normal phase is a second order phase transition
around the strong coupling limit.
On the other hand, in the weak coupling regime,
the Aoki phase is split into several cusps and shrinks to the poles corresponding to the doublers;
between these poles, the topology of the system remains non-trivial as in the free Hamiltonian.

\begin{figure}[tbp]
\begin{center}
\begin{tabular}{c}
\includegraphics[height=5cm]{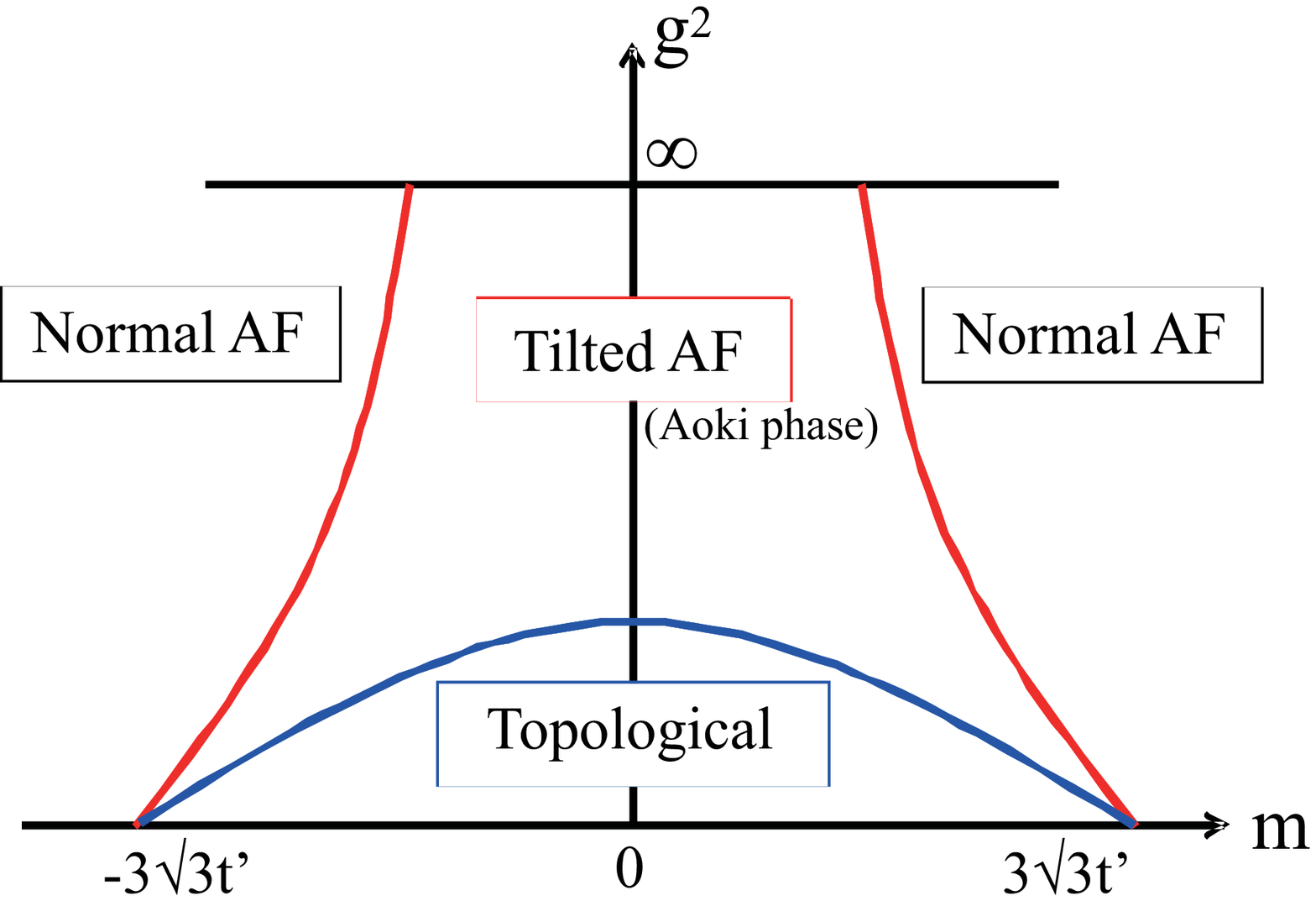} \\
(i) $t'<t'_C$ \\
\includegraphics[height=5cm]{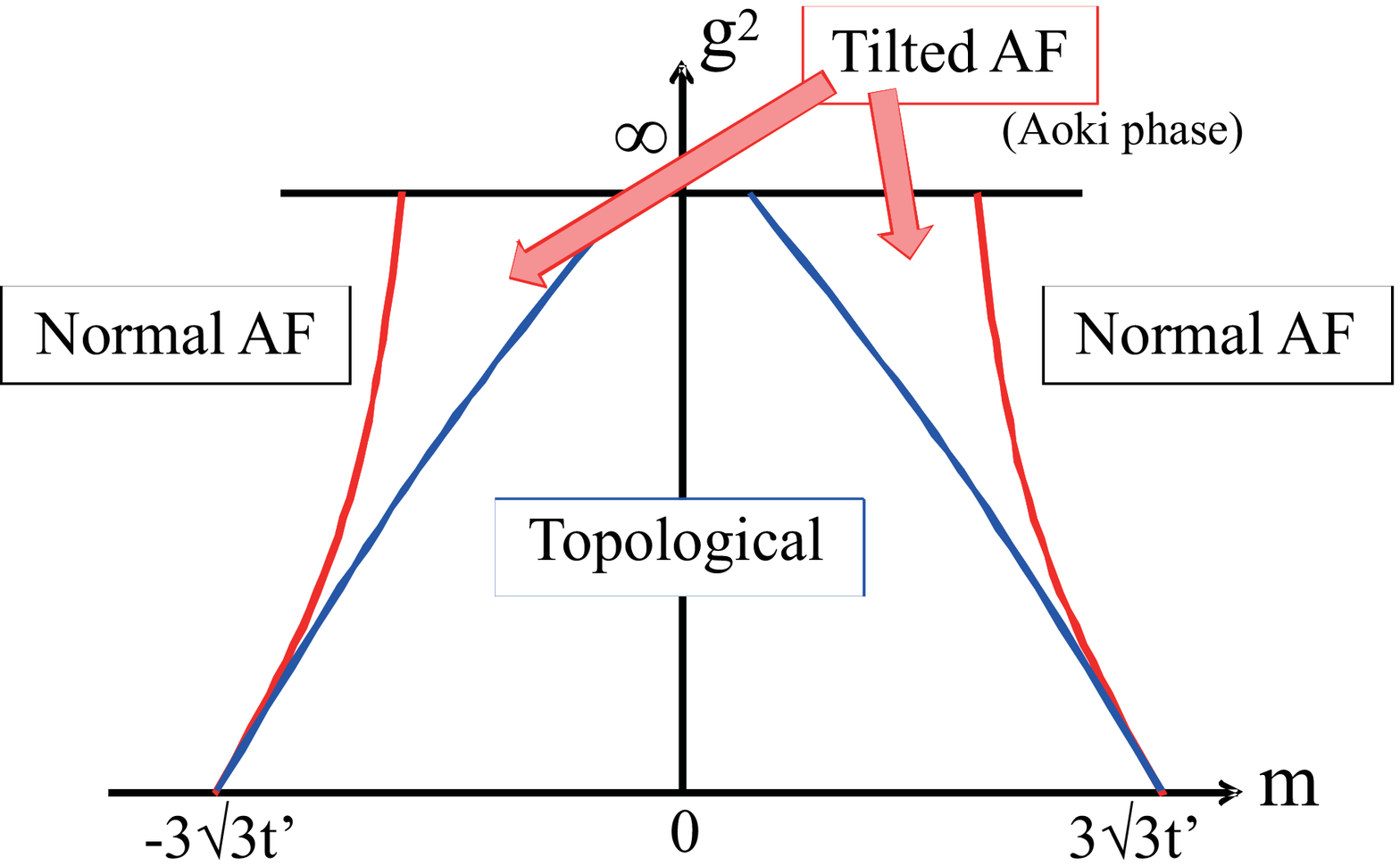} \\
(ii) $t'>t'_C$
\end{tabular}
\end{center}
\caption{The schematic phase structure of graphenelike system with the spin-orbit interaction,
conjectured in analogy to that of lattice QCD with the Wilson fermion formalism.
Here we fix the spin-orbit interaction $t'$,
and vary the electron-electron interaction strength $g^2$ and the uniform mass (staggered magnetic field) term $m$.
From the strong coupling limit $(g^2=\infty)$ investigated in our study,
the phase structure evolves to the weak coupling region differently,
depending on whether $t'$ is below or above the critical value $t'_C=0.0538t$.}
\label{fig:haldane-ph}
\end{figure}

Since the spin-orbit interaction on the honeycomb lattice and the Wilson term of lattice fermions
have the similar topological features as we have seen,
we can relate the ``tilted antiferromagnetic (AF)'' phase
observed in 2-dimensional topological insulators here to
the Aoki phase in the lattice QCD with the Wilson fermion with the following analogy (see Table \ref{table:wilson}).
The tilted AF phase is characterized by the nematic AF order $\langle \sigma_2 \rangle \neq 0$
in the U(1) remnant spin symmetry space,
while the Aoki phase is characterized by the pion condensation $\langle \bar{\psi} i\gamma_5 \psi \rangle \neq 0$
in the continuous chiral symmetry space.
In both cases, the emergent order parameter is orthogonal to the direction pointed by the external source,
namely the staggered magnetic field $S_M$ and the mass term $m \bar{\psi}\psi$ respectively.
Relying on this analogy,
we can make some conjecture about the phase structure of the graphenelike system
from strong coupling to weak coupling,
as shown schematically in Fig.\ref{fig:haldane-ph}:
\begin{itemize}
\item (i) When $t'<t'_C$, the system reveals the tilted AF (Aoki) phase around $m=0$ in the strong coupling limit.
Going down to the weak coupling region, this phase splits into two branches,
and shrinks to two points corresponding to the topological phase boundary in the free limit.
The topological (QSH) phase {with a finite spin Hall conductivity} appears
when the coupling strength reaches a sufficiently small value.
\item (ii) When $t'>t'_C$, the system reveals the topological phase around $m=0$ even in the strong coupling limit,
while the tilted AF phase appears at intermediate $m$.
Here the tilted AF phase is already split into two branches, corresponding to the sign of mass.
Going down to the weak coupling region,
the tilted AF phase shrinks to two points corresponding to the topological phase boundary.
\end{itemize}
Here we have assumed that the phase boundary is continuous from the strong coupling limit to the free regime.
The phase transition around the strong coupling limit is second-order as shown in our study,
while it remains unclear in the weak-coupling region whether it is first-order or second-order,
as it has been argued in lattice QCD \cite{Sharpe-Singleton}.

\section{Conclusion}\label{sec:summary}
In this paper,
we have observed the effect of the electron-electron interaction
on the phase structure of 2-dimensional topological insulators,
extending the idea of previous works with respect to the spontaneous symmetry breaking in graphene.
Kane-Mel\'e model on the honeycomb lattice is extended by introducing the effect of electron-electron interaction
mediated by electromagnetic field (U(1) gauge field).
By the techniques of strong coupling expansion of lattice gauge theory,
we have observed the behavior of the spontaneous antiferromagnetic (AF) order in the strong coupling limit of the interaction.
As a result, we have found that the topological phase structure is modified from that of the noninteracting system,
by the emergence of a new ``tilted AF'' phase in between the normal insulator and the topological insulator phases.
The AF order is not parallel to the direction pointed by the spin-orbit interaction and the staggered magnetic field
in the spin SU(2) space in this phase,
which will result in the emergence of a gapless Nambu--Goldstone mode corresponding to the in-plane rotation.
We have also shown the analogy between the phase structure of topological insulators shown here
and that of the strongly coupled lattice QCD with the Wilson fermion formalism.
The tilted AF phase is similar to the so-called ``Aoki phase'' in lattice QCD in that
both of them are characterized by an order parameter orthogonal to the external source term
in the continuous symmetry space.
Such an analogy may help us understand the behavior of topological insulators with an electron-electron interaction
from strong coupling to weak coupling regime.

There remain several open questions to be solved.
One is the restriction of the SU(2) spin symmetry space down to U(1),
due to the lattice discretization in the imaginary time direction.
Such a lattice artifact obscures the details of the NG mode appearing in the tilted AF phase,
so that its quantitative effect on the charge/spin transport properties is left to be calculated.
It also excludes the spin singlet orders,
such as a charge density wave and Haldane flux state, which induces quantum anomalous Hall effect \cite{Haldane,Raghu_2008,Weeks_2010}.
Such effects should be taken into account
by the models with an exact symmetry structure,
such as the extrapolation to a (hypothetical) multiflavor theory,
or by other techniques that do not require any lattice regularization process.
The relation to the physics in the realistic topological insulators would be another problem.
Comparison of our findings to the similar strong coupling analysis in the effective model of 3-dimensional topological insulators \cite{Sekine_2013},
such as Bi$_2$Se$_3$ or Bi$_2$Te$_3$,
would give us some clues.
The interaction effect on the quantum Hall states and the gapless surface (edge) states would be experimentally of a great importance.

\
      
\begin{acknowledgments}
The authors are thankful to T.~Hatsuda, K.~Nomura, T.~Oka, and N.~Tsuji for fruitful discussions.
Y.~A. is supported by Grant-in-Aid for Japan Society for the Promotion of Science (DC1, No.22.8037).
T.~K. is supported by Grant-in-Aid for Japan Society for the Promotion of Science (PD, No.23.593).
\end{acknowledgments}

\

\appendix

\section{Construction of lattice action} \label{app:path-integral}
Here we provide a rigorous description about the derivation of path integral formalism on the honeycomb lattice,
and give a physical interpretation to the doubling problem arising from the lattice discretization.
In the Hamiltonian formalism,
the dynamics of electrons and the electromagnetic field is given by the Hamiltonian
\begin{align}
H_F &= t \sum_{\bfr_A} \sum_{s=\pm} \sum_{i=1,2,3} s \left[a_s^\dag(\bfr_A) U_i(\bfr_A,0) b_s^\dag(\bfr_A+\bfs_i) + \mathrm{H.c.}\right] \\
H_G &= \frac{\sqrt{3} a^3}{2}  \sum_{z \in a \mathbb{Z}} \left\{ \sum_{\bfr_A, i}\left[E_i(\bfr_A,z)\right]^2 +\frac{3}{4}\sum_{\bfr \in A \cup B} \left[E_z(\bfr,z)\right]^2 \right\} \nonumber \\
 & \quad -\frac{1}{ae^2} \sum_{P} \left[P+P^\dag \right],
\end{align}
where the gauge field operator $A$ and its conjugate momentum (electric field) $E$ are defined on the lattice links as
\begin{align}
A_i(\bfr_A,z) & \equiv \frac{1}{a}\int_{\bfr_A}^{\bfr_A+\bfs_i} d\bfr\cdot \mathbf{A}(\bfr,z), \\
E_i(\bfr_A,z) & \equiv \frac{1}{a}\int_{\bfr_A}^{\bfr_A+\bfs_i} d\bfr\cdot \mathbf{E}(\bfr,z) \\
A_z(\bfr,z) & \equiv \frac{1}{a}\int_{z}^{z+a} dz' A(\bfr,z'), \\
E_z(\bfr,z) & \equiv \frac{1}{a}\int_{z}^{z+a} dz' E(\bfr,z'),
\end{align}
and $P$ is a plaquette operator constructed of link variables $U_i = \exp[iea A_i]$ and $U_z= \exp[iea A_z]$.
The summation $\sum_P$ is taken over all the plaquettes
(both honeycomb and square) on the lattice.
It should be noted that the gauge field operators are defined in the 3-dimensional space $(\bfr,z)$.
As for the fermions, we have performed a Bogoliubov transformation defined by
\begin{align}
a_\uparrow \rightarrow a_+, \quad a_\downarrow \rightarrow a_-^\dag, \quad b_\uparrow \rightarrow b_+^\dag, \quad b_\downarrow \rightarrow b_-,
\end{align}
where we should note that the labels $\uparrow$ and $\downarrow$ represent
the eigenvalue of an arbitrarily chosen spin operator.
Here we take it as the Pauli matrix $\sigma_y$ for later convenience.

By splitting the inverse temperature $\beta$ by an infinitesimal timeslice $\Delta\tau$ as
\begin{align}
Z = \mathrm{Tr} e^{-\beta H} = \mathrm{Tr} \left[e^{-\Delta\tau H} \cdots e^{-\Delta\tau H} \right]
\end{align}
and inserting complete sets of states between every $e^{-\Delta\tau H}$,
we obtain the Euclidean action on the honeycomb lattice,
\begin{align}
S_F = & \Delta\tau \left[ \sum_{\bfr_A,\tau}a_s^\dag \partial_\tau a_s + \sum_{\bfr_B,\tau}b_s^\dag \partial_\tau b_s \right] +\Delta\tau\sum_\tau H_F(\tau), \label{eq:s-f0} \\
S_G = & \frac{\sqrt{3} a^3 \Delta\tau}{2}  \left[ \sum_{\bfr_A, i, z, \tau}\left(\partial_\tau A_i \right)^2 +\frac{3}{4}\sum_{\bfr,z,\tau} \left(\partial_\tau A_z \right)^2 \right] \nonumber \\
& -\frac{\Delta\tau}{ae^2} \sum_{P,\tau} \left[P+P^\dag \right], \label{eq:s-g0}
\end{align}
where the derivative $\partial_\tau$ is defined as
$ \partial_\tau f(\tau) \equiv [f(\tau+\Delta\tau) - f(\tau)]/\Delta\tau $.
If we go back to the original spin representation, Eq.(\ref{eq:s-f0}) reads
\begin{align}
S_F =& \Delta\tau \sum_{\bfr_A,\tau} \left[ a_\sigma^\dag \partial_{\tau} a_\sigma + (\Delta\tau)(\partial_\tau a_\downarrow^\dag)(\partial_\tau a_\downarrow)\right] \\
& + \Delta\tau \sum_{\bfr_B,\tau} \left[ b_\sigma^\dag \partial_{\tau} b_\sigma + (\Delta\tau)(\partial_\tau b_\downarrow^\dag)(\partial_\tau b_\downarrow)\right] + \Delta\tau \sum_\tau H_F(\tau). \nonumber
\end{align}
Therefore, the lattice action does not preserve the global spin SU(2) symmetry
unless the continuum limit $\Delta\tau \rightarrow 0$ is taken,
but it is still invariant under the remnant U(1) rotation generated by $\sigma_y$.

Here we fix the lattice anisotropy $a/\Delta\tau \equiv \vf$,
to reproduce the ratio between the intrinsic cutoffs of energy and momentum given by the Dirac cone structure.
If we take the physical value in monolayer graphene, the ratio reads $\Delta\tau/a = \vf^{-1} \gg 1$,
so that we can apply saddle point approximation to the second line of Eq.(\ref{eq:s-g0}),
yielding
\begin{align}
P=1 \quad \mathrm{i.e.} \quad \nabla \times \mathbf{A}=0.
\end{align}
Thus we can take the scalar potential $\phi(\bfr,z,\tau)$,
which satisfies the relations
\begin{align}
U_i(\bfr_A,z,\tau) =& e^{iea A_i(\bfr_A,z,\tau)} = e^{i[\phi(\bfr_A+\bfs_i,z,\tau)-\phi(\bfr_A,z,\tau)]}, \\
U_z(\bfr,z,\tau) =& e^{iea A_z(\bfr,z,\tau)} = e^{i[\phi(\bfr,z+a,\tau)-\phi(\bfr,z,\tau)]}.
\end{align}
This approximation drops off the retardation of the electromagnetic field,
which is referred to as ``instantaneous approximation''.
Such an approximation enables us to reconstruct the gauge action only in terms of its temporal component.
By the local gauge transformation
\begin{equation}
a_s(\bfr,\tau) \rightarrow e^{-is \phi(\bfr,0,\tau)}a_s(\bfr,\tau), \; b_s(\bfr,\tau) \rightarrow e^{is \phi(\bfr,0,\tau)}b_s(\bfr,\tau)
\end{equation}
and taking a new link variable
\begin{align}
U_0(\bfr,z,\tau) \equiv e^{-i\left[\phi(\bfr,z,\tau+\Delta\tau)-\phi(\bfr,z,\tau)\right]} \equiv e^{-i\theta(\bfr,z,\tau)}, \label{eq:def-links}
\end{align}
we can set the spatial link variables to unity.
Due to this definition,
the Polyakov loop obeys the constraint
\begin{align}
\prod_{\tau} U_0(\bfr,z,\tau) =1, \quad \mathrm{i.e.} \quad \sum_{\tau} \theta(\bfr,z,\tau)=0. \label{eq:ploop}
\end{align}
As a result, the lattice action reads
\begin{align}
S_F =& \Delta\tau \sum_{\bfr_A,s,\tau} a_s^\dag(\bfr_A,\tau) \tfrac{[U_0(\bfr_A,\tau)]^s a_s(\bfr_A,0,\tau+\Delta\tau)-a_s(\bfr_A,\tau)}{\Delta \tau} \nonumber \\
 & + \Delta\tau \sum_{\bfr_B,s,\tau} b_s^\dag(\bfr_B,\tau) \tfrac{[U_0^\dag(\bfr_B,0,\tau)]^sb_s(\bfr_B,\tau+\Delta\tau)-b_s(\bfr_B,\tau)}{\Delta \tau} \nonumber \\
 & + \Delta\tau \sum_{\tau}H_F(\tau) \label{eq:lattice-s-f}\\
S_G =& \frac{\sqrt{3}}{2e^2}\frac{a}{\Delta\tau} \sum_{\bfr_A,i,z,\tau} \left[\theta(\bfr_A,z,\tau)-\theta(\bfr_A+\bfs_i,z,\tau)\right]^2 \nonumber \\
 & + \frac{3\sqrt{3}}{2e^2}\frac{a}{\Delta\tau} \sum_{\bfr \in A\cup B,z,\tau} \left[\theta(\bfr,z+a,\tau)-\theta(\bfr,z,\tau)\right]^2. \label{eq:lattice-s-g-nc}
\end{align}
The kinetic term of the gauge field in Eq.(\ref{eq:lattice-s-g-nc}) is given in the non-compact form.
In the continuum limit $(\Delta\tau\rightarrow 0)$,
it can be regularized by the compact form
\begin{align}
S_G =& -\sqrt{3}\beta \sum_{\bfr_A,i,z,\tau}\mathrm{Re} \left[ U_0(\bfr_A+\bfs_i,z,\tau) U_0^*(\bfr_A,z,\tau) \right] \nonumber \\
 & - 3\sqrt{3}\beta \sum_{\mathbf{r}\in A \cup B, z, \tau} \mathrm{Re} \left[U_0(\bfr,z+a,\tau) U_0^*(\bfr,z,\tau)\right]. \label{eq:lattice-s-g}
\end{align}

Let us prove that the fermionic action in Eq.(\ref{eq:lattice-s-f}) is equivalent
to the ``staggered fermion'' formalism, which has been naively given in Eq.(\ref{eq:s-f}).
Since $a_s$ and $b_s$ are not operators but just Grassmann variables,
we can propose the following change of integration variables in the path integral:
\begin{align}
a_+(\tau) \rightarrow \alpha(\tau), & \quad a_+^\dag(\tau) \rightarrow \bar{\alpha}(\tau+\Delta\tau'), \nonumber \\
a_-(\tau) \rightarrow \bar{\alpha}(\tau), & \quad a_-^\dag(\tau) \rightarrow \alpha(\tau+\Delta\tau') \nonumber \\
b_+(\tau) \rightarrow \bar{\beta}(\tau), & \quad b_+^\dag(\tau) \rightarrow \beta(\tau+\Delta\tau'), \nonumber \\
b_-(\tau) \rightarrow \beta(\tau), & \quad b_-^\dag(\tau) \rightarrow \bar{\beta}(\tau+\Delta\tau'), \label{eq:correspondence}
\end{align}
with a finer time-slice $\Delta\tau' \equiv \Delta\tau/2$.
The spin degrees of freedom are absorbed in the temporal lattice mesh.
The anti-periodicity in the temporal direction
\begin{align}
\alpha(\tau'+\beta)=-\alpha(\tau'), \quad \beta(\tau'+\beta) = -\beta(\tau')
\end{align}
also holds for the new fermionic fields.
By this transformation, the fermionic action reads
\begin{align}
S_F =& \sum_{\bfr_A,\tau'}\left[\bar{\alpha} V_0 \alpha' - \bar{\alpha}' V_0^\dag \alpha \right] + \sum_{\bfr_B,\tau'}\left[\bar{\beta} V_0 \beta' - \bar{\beta}' V_0^\dag \beta \right] \nonumber \\
& -2t\Delta\tau' \sum_{\bfr_A,i,\tau'} \left[\bar{\alpha} \beta + \bar{\beta} \alpha\right], \label{eq:s-f-stagg}
\end{align}
where $\chi' \equiv \chi(\tau'+\Delta\tau')$ for $\chi=\bar{\alpha},\bar{\beta},\alpha,\beta$.
The new link variable $V_0$ is defined by
\begin{align}
V_0(\bfr,\tau') \equiv \begin{cases} 1 & (\tau'/\Delta\tau'=\text{even}) \\ U_0(\bfr,\tau'-\Delta\tau') & (\tau'/\Delta\tau'=\text{odd}) \end{cases}.
\end{align}
This fermionic action is invariant under the global rotation corresponding to the remnant U(1) spin symmetry
defined in Eq.(\ref{eq:remnant}).
It agrees
with the staggered honeycomb lattice action in Eq.(\ref{eq:s-f}),
if we renormalize $\alpha$ and $\beta$ by the factor $1/\sqrt{2}$
and interpolate $V_{0,\mathrm{e}}$ by a dynamical U(1) link variable.
Therefore, the change of variables in Eq.(\ref{eq:correspondence}) gives
the correspondence between the true spin degrees of freedom and the staggered fermions on the honeycomb lattice.

\section{Absence of order parameter $\sigma_1$ at $m=0$} \label{app:sigma=0}
In this appendix, we show how the order parameter $\sigma_1$ gets suppressed
in the absence of the staggered magnetic field $m$,
by solving the gap equations (\ref{eq:gapeq-1}) and (\ref{eq:gapeq-2}).

Let us assume the solution $\tilde{\sigma}_1 \neq 0$.
Since the potential $F_\mathrm{eff}$ is even both in $\sigma_1$ and $\sigma_2$ at $m=0$,
we can set $\tilde{\sigma}_1 >0$ without losing generality.
The solution of Eq.(\ref{eq:gapeq-2}) is twofold:

\textbf{(i)}
If $\tilde{\sigma}_2=0$,
the convexity around $(\tilde{\sigma}_1,\tilde{\sigma}_2)$ in $\sigma_2$-direction reads
\begin{align}
\frac{\partial^2 F_\mathrm{eff}}{\partial \sigma_2^2} \Bigr|_{(\tilde{\sigma}_1,0)} &= 1 - \int_{\Omega}d^2\bfk \frac{1/2}{\left[\mathcal{E}(\tilde{\sigma}_1,0,t';\bfk)\right]^2} \label{eq:haldane-gap1} \\
 &= \frac{1}{\tilde{\sigma}_1} \left[\tilde{\sigma}_1 - \int_{\Omega}d^2\bfk \frac{\tilde{\sigma}_1/2}{\left[\mathcal{E}(\tilde{\sigma}_1,0,t';\bfk)\right]^2}\right] \nonumber \\
 &= -\frac{1}{\tilde{\sigma}_1} \int_{\bfk\in\Omega}d^2\bfk \frac{2t' \im \Phi_2(\bfk)}{\left[\mathcal{E}(\tilde{\sigma}_1,0,t';\bfk)\right]^2}, \nonumber
\end{align}
where we have used Eq.(\ref{eq:gapeq-1}).
Since $\Phi(-\bfk)=\Phi^*(\bfk)$ and $\Phi_2(-\bfk)=\Phi^*(\bfk)$,
we can separate the Brillouin zone $\Omega$ into two regions $\Lambda_\pm$ corresponding to the sign of $2t'\im\Phi_2(\bfk)$,
which yields
\begin{align}
& \frac{\partial^2 F_\mathrm{eff}}{\partial \sigma_2^2} \Bigr|_{(\tilde{\sigma}_1,0)} = -\frac{1}{\tilde{\sigma}_1} \sum_{\pm} \int_{\Lambda_\pm} d^2\bfk \frac{2t' \im \Phi_2(\bfk)}{\left[\mathcal{E}(\tilde{\sigma}_1,0,t';\bfk)\right]^2}\\
& = -\frac{1}{\tilde{\sigma}_1} \int_{\Lambda_+} d^2\bfk \left\{ \frac{2t' \im \Phi_2(\bfk)}{\left[\mathcal{E}(\tilde{\sigma}_1,0,t';\bfk)\right]^2} + \frac{-2t' \im \Phi_2(\bfk)}{\left[\mathcal{E}(\tilde{\sigma}_1,0,-t';\bfk)\right]^2 } \right\}. \nonumber
\end{align}
Since $\tilde{\sigma}_1>0$ and $2t'\im\Phi_2(\bfk)>0$ in $\Lambda_+$,
$\left[\mathcal{E}(\tilde{\sigma}_1,0,t';\bfk)\right]^2 = (\tilde{\sigma}_1/2-2t'\im\Phi_2(\bfk))^2+|t\Phi(\bfk)|^2$
is smaller than $\left[\mathcal{E}(\tilde{\sigma}_1,0,-t';\bfk)\right]^2 = (\tilde{\sigma}_1/2+2t'\im\Phi_2(\bfk))^2+|t\Phi(\bfk)|^2$.
Therefore, the convexity $\partial^2 F_\mathrm{eff}/\partial \sigma_2^2$ becomes negative,
which disagrees with the assumption that $(\tilde{\sigma}_1,\tilde{\sigma}_2)$ is the minimum.

\textbf{(ii)}
If $(1/\tilde{\sigma}_2)(\partial F_\mathrm{eff}/\partial \sigma_2)|_{(\tilde{\sigma}_1,\tilde{\sigma}_2)}=0$,
we have
\begin{align}
1 - \int_{\Omega}d^2\bfk \frac{1/2}{\left[\mathcal{E}(\tilde{\sigma}_1,\tilde{\sigma}_2,t';\bfk)\right]^2}=0.
\end{align}
Using this relation, the derivative in the $\sigma_1$-direction becomes
\begin{align}
\frac{\partial F_\mathrm{eff}}{\partial \sigma_1}\Bigr|_{(\tilde{\sigma}_1,\tilde{\sigma}_2)} &= \tilde{\sigma}_1 - \int_{\Omega}d^2\bfk \frac{\tilde{\sigma}_1/2-2t'\im\Phi_2(\bfk)}{\left[\mathcal{E}(\tilde{\sigma}_1,\tilde{\sigma}_2,t';\bfk)\right]^2} \\
 & = \int_{\Omega}d^2\bfk \frac{2t'\im\Phi_2(\bfk)}{\left[\mathcal{E}(\tilde{\sigma}_1,\tilde{\sigma}_2,t';\bfk)\right]^2}, \nonumber
\end{align}
which becomes nonzero unless $\tilde{\sigma}_1=0$,
as shown in the case (i).
It disagrees with the gap equation in Eq.(\ref{eq:gapeq-1}).

Therefore, we can conclude that the assumption $\tilde{\sigma}\neq 0$ is incorrect,
i.e. the order parameter $\tilde{\sigma}$ is completely tilted to the $\sigma_2$-direction.

\section{Evolution of the order parameter $\sigma$} \label{app:evolution}
In this appendix, we investigate how the order parameter $(\tilde{\sigma}_1,\tilde{\sigma}_2)$ evolves
as a function of $t'$ and $m$ in detail,
and discuss how the modified topological phase structure is related with that of the noninteracting system.

Here we fix the spin-orbit coupling amplitude $t'$ and vary the staggered magnetic field $m$.
In order to find the potential minimum in the $(\sigma_1,\sigma_2)$-plane,
first we fix $\sigma_1$ and check the sign of 
\begin{align}
\frac{\partial F_\mathrm{eff}}{\partial (\sigma_2^2)} = 1 - \int_{\Omega}d^2\bfk \frac{1/2}{\left[\mathcal{E}({\sigma}_1,{\sigma}_2,t';\bfk)\right]^2} \label{eq:haldane-s2}
\end{align}
instead of $\partial F_\mathrm{eff}/\partial \sigma_2$
(because $F_\mathrm{eff}(\sigma_1,\sigma_2)$ is always even in $\sigma_2$).
Since the right hand side of Eq.(\ref{eq:haldane-s2}) monotonically increases
and asymptotically reaches toward unity as a function of $\sigma_2(>0)$,
we have to consider two cases depending on its sign at $\sigma_2=0$:
\begin{itemize}
\item \textbf{(a)} If $\partial F_\mathrm{eff}/\partial (\sigma_2^2) |_{\sigma_2=0} \geq 0$,
the effective potential is minimized at $\tilde{\sigma}_2=0$.
\item \textbf{(b)} If $\partial F_\mathrm{eff}/\partial (\sigma_2^2) |_{\sigma_2=0} < 0$,
the effective potential is minimized at finite $\tilde{\sigma}_2$,
which satisfies $\partial F_\mathrm{eff}/\partial (\sigma_2^2)|_{\tilde{\sigma}_2}=0$.
Here we fix $\tilde{\sigma}_2$ to be positive.
\end{itemize}
We can regard $\tilde{\sigma}_2$ as a function of $\sigma_1$.
The curve composed of the set of points $\{(\sigma_1,\tilde{\sigma}_2(\sigma_1)) | \sigma_1 \geq 0 \}$ in the $(\sigma_1,\sigma_2)$-plane,
which we call here $C(t')$, is uniquely determined by the parameter $t'$,
and continuous in $\sigma_1$ because of the analyticity of Eq.(\ref{eq:haldane-s2}).


Along this curve $C(t')$, the overall potential minimum shall be found by varying $\sigma_1$.
By solving the equation
\begin{align}
& \frac{\partial F_\mathrm{eff}}{\partial \sigma_1} \Bigr|_{(\tilde{\sigma}_1,\tilde{\sigma}_2(\tilde{\sigma}_1))} \left( \equiv f(\tilde{\sigma}_1) \right) \label{eq:haldane-solve} \\
&= \tilde{\sigma}_1-2m - \int_{\Omega}d^2\bfk \frac{\tilde{\sigma}_1/2-2t'\im\Phi_2(\bfk)}{\left[\mathcal{E}(\tilde{\sigma}_1,\tilde{\sigma}_2(\tilde{\sigma}_1),t';\bfk)\right]^2} =0, \nonumber
\end{align}
we obtain the potential minimum $(\tilde{\sigma}_1,\tilde{\sigma}_2(\tilde{\sigma}_1))$,
as a function of the parameters $t'$ and $m$.

First, we show that there is a unique one-to-one correspondence between $\tilde{\sigma}_1$ and $m$,
when $t'$ is fixed to a finite value.
We can easily see that $\tilde{\sigma}_1(m=0)=0.$
On the other hand, $f(\tilde{\sigma}_1)$ asymptotically becomes $\tilde{\sigma}_1-2m$
as $\tilde{\sigma}_1 \rightarrow \infty$ (note that $\tilde{\sigma}_2=0$ in this limit),
so that we obtain the asymptotic solution $\tilde{\sigma}_1(m\rightarrow\infty) \sim 2m$.

Now that the boundaries of $m$ and those of $\tilde{\sigma}_1$ are matched,
we check whether $\tilde{\sigma}_1$ monotonically increases between these boundaries as a function of $m$ or not.
By differentiating both sides of Eq.(\ref{eq:haldane-solve}) by $m$,
we have the relation
\begin{align}
\frac{\partial \tilde{\sigma}_1}{\partial m} f'(\tilde{\sigma}_1) =2.
\end{align}
Thus, what we have to show is that the factor $f'(\tilde{\sigma}_1) \geq 0$ for any value of $\tilde{\sigma}_1(>0)$.
Here we consider again the regions (a) and (b) given above:
\begin{itemize}
\item \textbf{(a)} In the region where $\tilde{\sigma}_2(\tilde{\sigma}_1)=0$,
we have the relation
\begin{align}
\frac{\partial F_\mathrm{eff}}{\partial (\sigma_2^2)} \Bigr|_{(\tilde{\sigma}_1,0)} =
1 - \int_{\Omega}d^2\bfk \frac{1/2}{\left[\mathcal{E}(\tilde{\sigma}_1,0,t';\bfk)\right]^2} \geq 0,
\end{align}
from the definition of this region.
This relation yields
\begin{align}
f'(\tilde{\sigma}_1) = & 1-\int_{\Omega}d^2\bfk \frac{1/2}{\left[\mathcal{E}(\tilde{\sigma}_1,0,t';\bfk)\right]^2} \\
 & \quad + \int_{\Omega}d^2\bfk \frac{\left[\tilde{\sigma}_1/2-2t'\im\Phi_2(\bfk) \right]^2}{\left[\mathcal{E}(\tilde{\sigma}_1,0,t';\bfk)\right]^4} \geq 0. \nonumber
\end{align}
\item \textbf{(b)} In the region where $\tilde{\sigma}_2(\tilde{\sigma}_1) \neq 0$,
we have the relation
\begin{align}
\frac{\partial F_\mathrm{eff}}{\partial (\sigma_2^2)}\Bigr|_{(\tilde{\sigma}_1,\tilde{\sigma}_2)}
 = 1 - \int_{\Omega}d^2\bfk \frac{1/2}{\left[\mathcal{E}(\tilde{\sigma}_1,\tilde{\sigma}_2(\tilde{\sigma}_1),t';\bfk)\right]^2} =0.
\end{align}
By differentiating both sides of this relation by $\tilde{\sigma}_1$, we obtain a new relation
\begin{align}
\int_{\Omega}d^2\bfk \frac{\tilde{\sigma}_1/2-2t'\im\Phi_2(\bfk)+\tilde{\sigma}'_2 \tilde{\sigma}_2 /2}{\left[\mathcal{E}(\tilde{\sigma}_1,\tilde{\sigma}_2(\tilde{\sigma}_1),t';\bfk)\right]^4} =0,
\end{align}
where $\tilde{\sigma}'_2 \equiv \partial \tilde{\sigma}_2(\tilde{\sigma}_1) / \partial \tilde{\sigma}_1$.
Using these two relations, we can simplify $f'(\tilde{\sigma}_1)$ as
\begin{align}
& f'(\tilde{\sigma}_1) = 1 - \int_{\Omega}d^2\bfk \frac{1/2}{\left[\mathcal{E}(\tilde{\sigma}_1,\tilde{\sigma}_2(\tilde{\sigma}_1),t';\bfk)\right]^2} \nonumber \\
& \quad +\int_{\Omega}d^2\bfk \frac{\left[\tfrac{\tilde{\sigma}_1}{2}-2t'\im\Phi_2(\bfk) \right] \left[\tfrac{\tilde{\sigma}_1}{2}-2t'\im\Phi_2(\bfk)+\tilde{\sigma}'_2 \tilde{\sigma}_2 /2 \right]}{\left[\mathcal{E}(\tilde{\sigma}_1,\tilde{\sigma}_2(\tilde{\sigma}_1),t';\bfk)\right]^4} \nonumber \\
&= \int_{\bfk} d^2\bfk \frac{ \left[ \tilde{\sigma}_1/2-2t'\im\Phi_2(\bfk)+\tilde{\sigma}'_2 \tilde{\sigma}_2 /2 \right]^2}{\left[\mathcal{E}(\tilde{\sigma}_1,\tilde{\sigma}_2(\tilde{\sigma}_1),t';\bfk)\right]^4} \geq 0.
\end{align}
\end{itemize}
Therefore, $\tilde{\sigma}_1(m)$ monotonically increases for any $m \in [0,\infty)$.

Due to this one-to-one correspondence between $m$ and $\tilde{\sigma}_1$,
the solution $(\tilde{\sigma}_1,\tilde{\sigma}_2)$ moves continuously along the path $C(t')$,
starting from $\tilde{\sigma}_1(m=0)=0$ toward $\tilde{\sigma}_1\rightarrow\infty$
as shown in Fig.\ref{fig:haldane-path},
when $t'$ is fixed and $m$ is varied.
Thus, as shown in Fig.\ref{fig:haldane-s12},
the order parameter $\sigma_1$ monotonically increases as a function of $m$,
while $\sigma_2$ shows a transition between $\sigma_2 \neq 0$ and $\sigma_2 =0$.
From here on,
we employ the ``modified'' effective mass $\tilde{\sigma}_1/2$ as the parameter characterizing the system,
instead of the bare mass $m$,
to discuss the phase transition characterized by $\sigma_2$.
(When $\tilde{\sigma}_1$ is given, we can derive the value of $m$ by Eq.(\ref{eq:haldane-solve}).)

\begin{figure}[tbp]
\begin{center}
\includegraphics[width=8.5cm]{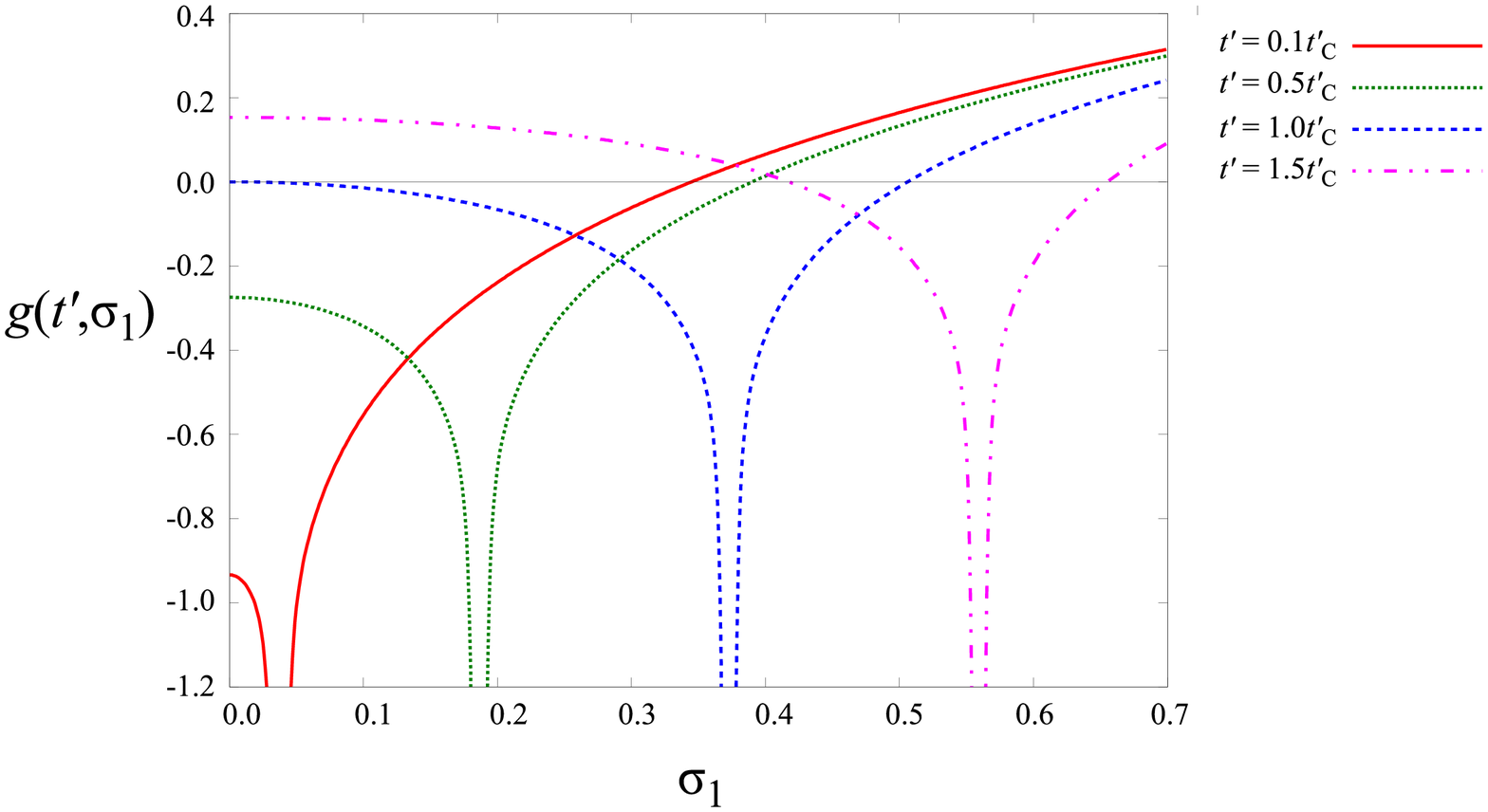}
\end{center}
\caption{The behavior of the function $g(t',\sigma_1)$ for several values of $t'$.
This function shows logarithmic divergence at $\sigma_1/2=3\sqrt{3}t'$,
where one of the valleys totally loses its spectral gap.
}
\label{fig:lhs}
\end{figure}

The phase structure of the system is related to the behavior of the curve $C(t')$.
The curve leaves from the $\sigma_1$-axis, namely $\tilde{\sigma}_2 \neq 0$, in the region (b),
which corresponds to the ``tilted antiferromagnetic'' phase shown in Fig.\ref{fig:haldane-phasediagram}.
On the other hand, the curve coincides with the $\sigma_1$-axis in the region (a),
which can be classified into the conventional or topological insulator phases.
According to the definition of the regions (a) and (b),
the phase structure of the system is characterized by the sign of the factor
\begin{align}
g(t',\tilde{\sigma}_1) \equiv  \frac{\partial F_\mathrm{eff}}{\partial (\sigma_2^2)}\Bigr|_{(\tilde{\sigma}_1,0)} = 1 - \int_{\Omega}d^2\bfk \frac{1/2}{\left[\mathcal{E}(\tilde{\sigma}_1,0,t';\bfk)\right]^2}.
\end{align}
The behavior of $g(t',\tilde{\sigma}_1)$ for several values of $t'$ is shown in Fig.\ref{fig:lhs}.
This function shows a negative logarithmic divergence at $\tilde{\sigma}/2 =3\sqrt{3}t'$ for any value of $t'(\neq 0)$,
since one of the valleys becomes gapless at this point.
Therefore, the system under an electron-electron interaction shows the tilted AF phase
around the topological phase boundary originally given in the noninteracting system.
When $t'$ and $m$ become dominant compared to the electron-electron interaction,
two phase boundaries around the tilted AF phase approach asymptotically to the original topological phase boundary.


\end{document}